\documentclass[11pt,a4paper]{article}
\usepackage{pslatex}
\usepackage{amsfonts}
\usepackage{amssymb}
\usepackage{amsmath}
\usepackage{eucal}
\usepackage{accents}
\usepackage{graphics}                 
\usepackage{color}                    
\usepackage{hyperref}                 
\usepackage{fancybox} 	
\usepackage{texdraw}
\usepackage{eucal}
\usepackage{epic}
\usepackage{epsfig}
\usepackage{picinpar}
\def\m@th{\mathsurround=0pt}
\mathchardef\bracell="0365
\def\upbrall{$\m@th\bracell$}
\def\undertilde#1{\mathop{\vtop{\ialign{##\crcr
ššš $\hfil\displaystyle{#1}\hfil$\crcr
šššš \noalign
šššš {\kern1.5pt\nointerlineskip}
šššš \upbrall\crcr\noalign{\kern1pt
šš }}}}\limits}

\mathchardef\braceup"0371
\def\upbroll{$\m@th\braceup$}
\def\underhat#1{\mathop{\vtop{\ialign{##\crcr
ššš $\hfil\displaystyle{\widehat{#1}}\hfil$\crcr
ššš $\hfil\displaystyle{#1}\hfil$\crcr
šššš \noalign
šššš {\kern1.5pt\nointerlineskip}
šššš \upbrall\crcr\noalign{\kern1pt
šš }}}}\limits}
\def\m@th{\mathsurround=0pt}
\mathchardef\bracell="0365 
\def\upbrall{$\m@th\bracell$}
\def\undertilde#1{\mathop{\vtop{\ialign{##\crcr
    $\hfil\displaystyle{#1}\hfil$\crcr
     \noalign
     {\kern1.5pt\nointerlineskip}
     \upbrall\crcr\noalign{\kern1pt
   }}}}\limits}

\mathchardef\braceup"0371 
\def\upbroll{$\m@th\braceup$}
\def\underhat#1{\mathop{\vtop{\ialign{##\crcr
    $\hfil\displaystyle{\widehat{#1}}\hfil$\crcr
    $\hfil\displaystyle{#1}\hfil$\crcr
     \noalign
     {\kern1.5pt\nointerlineskip}
     \upbrall\crcr\noalign{\kern1pt
   }}}}\limits}

\newcommand{\en}{\epsilon}

\newcommand{\sg}{\sigma}

\newcommand{\be}{\begin{equation}}
\newcommand{\ee}{\end{equation}}
\newcommand{\beq}{\begin{displaymath}}
\newcommand{\eeq}{\end{displaymath}}
\newcommand{\bee}{\begin{displaymath}}
\newcommand{\eee}{\end{displaymath}}
\newcommand{\bea}{\begin{eqnarray}}
\newcommand{\eea}{\end{eqnarray}}
\newcommand{\beaq}{\begin{eqnarray*}}
\newcommand{\eeaq}{\end{eqnarray*}}
\newcommand{\bse}{\begin{subequations}}
\newcommand{\ese}{\end{subequations}}

\newcommand{\wt}{\widetilde}
\newcommand{\wb}{\overline}

\newcommand{\ab}{\alpha,\beta}
\newcommand{\ba}{\beta,\alpha}

\definecolor{red}{rgb}{1,0,0}           
  \definecolor{green}{rgb}{0,1,0}
  \definecolor{blue}{rgb}{0,0,1}
  
\definecolor{Light}{gray}{.80}          
\definecolor{Dark}{gray}{.20}
\definecolor{pink}{rgb}{.95,0.82,0.92}  

\author{Chris M Field
\\
\texttt{cmfield@maths.usyd.edu.au}
\\
\noindent
School of Mathematics and Statistics F07, The University of Sydney, Sydney, \\Australia
}

\title{Extension of the Adler-Bobenko-Suris classification of integrable lattice equations}

\begin{document}

\maketitle

\begin{abstract}

\noindent

The classification of lattice equations that are integrable 
in the sense of higher-dimensional 
consistency is extended by allowing directed edges.
We find two cases that are 
not transformable via the `admissible transformations' 
to the lattice equations in the existing classification.

\end{abstract}

PACS: 02.30.Ik, 04.60.Nc, 05.45.Yv

\pagestyle{headings}

\section{Introduction}

Integrable systems come in many forms.  It can be argued that of these different 
forms (which include partial and ordinary differential equations, differential-difference 
equations, and difference equations) lattice (or `partial difference') equations 
are the most fundamental: they include not just one continuous integrable equation 
in the continuum limit, but entire hierarchies (see \cite{Miwa:hir}, \cite{WiCa:lattice}).  For 
systems defined on the vertices of a two-dimensional quadrilateral lattice 
or, more generally, on quad-graphs, the definition of integrability is so natural 
and transparent that it adds further weight to the argument for the primacy of 
such systems in the pantheon of integrability.  For these systems integrability 
is synonymous with higher-dimensional consistency.  Integrable lattice equations
are, in a sense, their own Lax pair (or `zero-curvature condition') 
-- the Lax pair follows in an algorithmic manner from the lattice equation (and 
the lattice equation follows from the Lax pair).  The clarification of three-dimensional 
consistency as the definition of integrability for such systems and the exploitation 
of the definition for the derivation of the Lax pair is a recent development, 
\cite{NiWa:Disc}, \cite{BobSu:Intquad}, \cite{Ni:Adl} 
(we refer to these papers for further details).

The importance of such systems established, 
it becomes desirable to have a classification, and a notion 
of which transformations can be performed on a lattice equation such that it 
still belongs to the same position in the classification.  
These issues were confronted for the first time in the landmark paper \cite{AdBoSu:Class}.

In \cite{AdBoSu:Class}, Adler, Bobenko and Suris (``ABS'') classified `integrable' equations on
quadrilateral lattices under certain symmetry assumptions 
(where integrability is equated with three-dimensional consistency) 
and specified the admissible transformations that may be performed on these equations such that they are still considered to be the same system (that is, take the same place in the classification).

The ABS classification has already received further study: reductions of the `Q' list to 
second-order mappings was performed by Joshi \textit{et al.} 
in \cite{Josh:nonQRT}, 
in \cite{Hiet:search} Hietarinta studied the necessity of the `tetrahedron property' 
assumption,
and symmetries of the lattice equations 
in the classification were investigated by Rasin and Hydon in \cite{Ras:symm}.  The classification has influenced other papers to varying degrees.

In the present paper a symmetry assumption of ABS is generalized in a natural way.  
This \textbf{extends the classification by the addition of the two integrable lattice equations}

\bee
Q = (x-y)(v-u) + (\alpha+\beta)(u-v) + (\beta-\alpha) (x-y) + \lambda(\alpha)-\lambda(\beta), 
\eee
\bee
Q = -\mu(-\beta) \, xu - \mu(\beta) \, vy  + \mu(-\alpha) \, xv + \mu(\alpha) \, uy , 
\eee
where $\lambda$ is any even function, and $\mu$ is any function 
that is not even.  We call these equations B1 and B2 respectively.
They are inequivalent to the existing members of the classification 
(they cannot be transformed to existing members of the classification 
by `admissible transformations').  The result here is that a 
generalization of one of the symmetry assumptions results in a new 
classification, consisting of the lattice equations in the original ABS 
classification, plus the 
addition of two lattice equations that obey the new symmetry (but not the 
more restrictive form stipulated by ABS).

In section \ref{ABSassumpt} 
the assumptions made by ABS in \cite{AdBoSu:Class} are described in detail.  
In section \ref{AsumptChang} the change to the symmetry assumptions is given; 
the modification to the derivation of ABS due to this change is given in section \ref{Der:a}.  
In section \ref{der:c} 
the new lattice equations are derived, 
and they are discussed in section \ref{discussion}.

\section{A review of the ABS assumptions and derivation}\label{ABSassumpt}

The equations on quadrilaterals considered by ABS are of the type

	\be\label{latticeeqn}
		Q(x, x_i, x_j, x_{i j}; \alpha_i, \alpha_j) = 0,
	\ee
where the field variables $x, x_i, x_j, x_{i j} \in \mathbb{C}$
are attached to the vertices of the quadrilateral (subscripts
$i$ and $j$ denoting steps along the lattice in directions $i$
and $j$ respectively) and `lattice parameters' $\alpha_i$
and $\alpha_j$ are associated to edges running in the $i$
and $j$ directions respectively (see fig. 1).

\begin{center}
\setlength{\unitlength}{0.001in}
\begingroup\makeatletter\ifx\SetFigFont\undefined%
\gdef\SetFigFont#1#2#3#4#5{%
  \reset@font\fontsize{#1}{#2pt}%
  \fontfamily{#3}\fontseries{#4}\fontshape{#5}%
  \selectfont}%
\fi\endgroup%
{\renewcommand{\dashlinestretch}{30}
\begin{picture}(2700,2066)(0,0)

\put(450,1883){\circle{150}}
\put(75,1883){\makebox(0,0)[lb]{$x_j$}} 

\put(2250,83){\circle{150}}
\put(2550,8){\makebox(0,0)[lb]{$x_{i}$}} 

\put(2250,1883){\circle{150}}
\put(2550,1883){\makebox(0,0)[lb]{$x_{ij}$}} 

\put(450,83){\circle{150}}
\put(75,8){\makebox(0,0)[lb]{$x$}} 

\put(1312.5,-75){\makebox(0,0)[lb]{$\alpha_i$}} 
\put(75,945.5){\makebox(0,0)[lb]{$\alpha_j$}} 

\drawline(450,1883)(450,83)
\drawline(450,1883)(2250,1883)
\drawline(2250,1883)(2250,83)
\drawline(450,83)(2250,83)
\end{picture}
}
\\

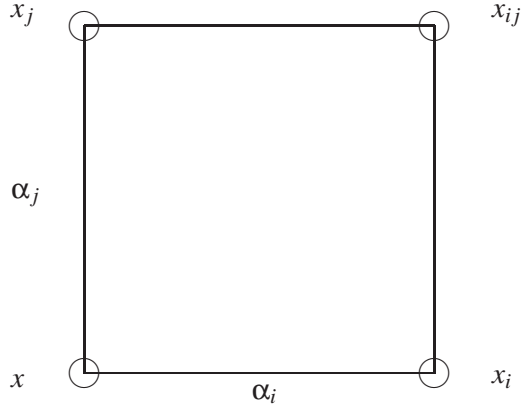
Fig 1.  An elementary quadrilateral.  
\end{center}

Lattice equations of the form (\ref{latticeeqn}) are classed as
integrable if they have the property of three-dimensional consistency.
To understand what `three-dimensional consistency' means,
consider the lattice equation (\ref{latticeeqn}) to be defined on the faces 
of a three-dimensional combinatorial cube, where parallel edges carry the same lattice parameter.
For instance, labelling the three orthogonal directions $(1,2,3)$, edges 
in the $1$, $2$, and $3$-direction 
have the lattice parameter $\alpha_1$, $\alpha_2$, and $\alpha_3$ attached, respectively.
Considering the initial data $x, x_1, x_2, x_3$ to be given, three-dimensional
consistency states that the three different ways of deducing $x_{1 2 3}$ 
all give the same result (see fig. 2).

\begin{center}
\setlength{\unitlength}{0.001in}
\begingroup\makeatletter\ifx\SetFigFont\undefined%
\gdef\SetFigFont#1#2#3#4#5{%
  \reset@font\fontsize{#1}{#2pt}%
  \fontfamily{#3}\fontseries{#4}\fontshape{#5}%
  \selectfont}%
\fi\endgroup%
{\renewcommand{\dashlinestretch}{30}
\begin{picture}(3482,2813)(0,-10)

\put(450,1883){\circle*{150}}
\put(75,1883){\makebox(0,0)[lb]{$x_3$}} 

\put(1275,2708){\circle{150}}
\put(825,2708){\makebox(0,0)[lb]{$x_{23}$}} 

\put(3075,2708){\circle{150}}
\put(3375,2633){\makebox(0,0)[lb]{$x_{123}$}} 

\put(2250,83){\circle*{150}}
\put(2550,8){\makebox(0,0)[lb]{$x_{1}$}} 

\put(1275,908){\circle*{150}}
\put(825,908){\makebox(0,0)[lb]{$x_2$}} 

\put(2250,1883){\circle{150}}
\put(1950,2108){\makebox(0,0)[lb]{$x_{13}$}} 

\put(450,83){\circle*{150}}
\put(0,8){\makebox(0,0)[lb]{$x$}} 

\put(3075,908){\circle{150}}
\put(3300,833){\makebox(0,0)[lb]{$x_{12}$}} 

\drawline(1275,2708)(3075,2708)
\drawline(1275,2708)(450,1883)
\drawline(450,1883)(450,83)
\drawline(3075,2708)(2250,1883)
\drawline(450,1883)(2250,1883)
\drawline(3075,2633)(3075,908)
\dashline{40.000}(1275,908)(450,83)
\dashline{40.000}(1275,908)(3075,908)
\drawline(2250,1883)(2250,83)
\drawline(450,83)(2250,83)
\drawline(3075,908)(2250,83)
\dashline{40.000}(1275,2633)(1275,908)
\end{picture}
}
\\

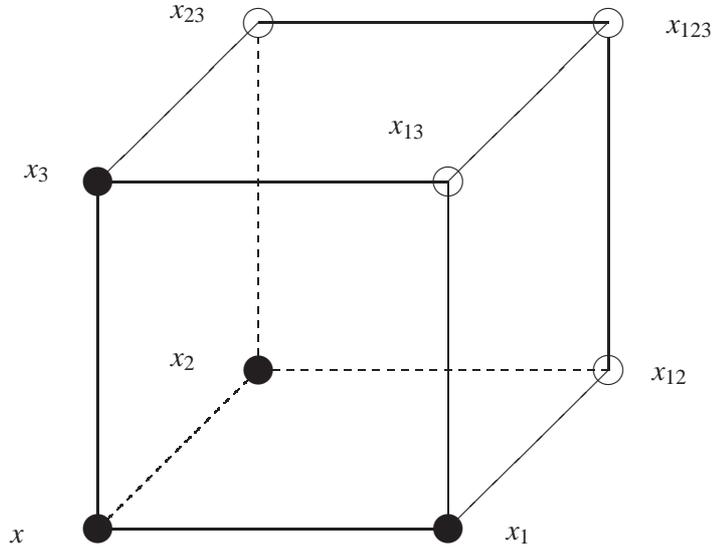
Fig 2. Three-dimensional consistency.  Given initial values at the black dots, $x, x_1, x_2, x_3$, 
the three possible ways to calculate $x_{123}$ all give the same result. 
\end{center}

For notational reasons, when considering only one plane (say the $(1,2)$ plane) we will denote $\alpha_1 \equiv \alpha$, $\alpha_2 \equiv \beta$, $x_1 \equiv u$, $x_2 \equiv v$, $x_{12} \equiv y$.

The assumptions under which ABS solved the classification problem of such equations are listed below.

\textbf{ 1] Linearity.}
\\
$Q(x, u, v, y; \alpha, \beta)$ is affine linear in each argument:

	\beq
		Q(x,u,v,y;\alpha,\beta) = a_1 x u v y + \ldots + a_{16}
	\eeq
where $\{a_i\}$ depend on $\alpha, \beta$.

\textbf{ 2] Symmetry.}
\\
$Q(x, u, v, y; \alpha, \beta)$ is invariant under the group of square symmetries:

	\be\label{sym2}
		Q(x,u,v,y;\alpha,\beta) = \epsilon Q(x,v,u,y;\beta,\alpha) = \sigma Q(u,x,y,v;\alpha,\beta)
	\ee
$\epsilon, \sigma = \pm 1$.

\textbf{ 3] Tetrahedron property.}
\\
The function 
	\begin{displaymath}
x_{123} = z (x,x_1,x_2,x_3;\alpha_1,\alpha_2,\alpha_3),
	\end{displaymath}
obtained from three-dimensional consistency,
does not depend on the variable $x$, that is, $z_x = 0$.

The following transformations are assumed to identify equivalence classes:
\begin{itemize}

	\item Action on all field variables $\{x_i\}$ by one and the same (independent of lattice parameter) M\"obius transformation.
	
	\item Simultaneous point change of all parameters $\alpha \mapsto \phi(\alpha)$.

\end{itemize}
These transformations do not violate the three assumptions listed above.

That these are the only admissible transformations is a key point of the classification.
It is stated in \cite{AdBoSu:Class} that the `A' list is obtainable from the `Q'
list, and the lower `Q' and `H' lattice equations (Q1 to Q3, and H1 to H2, respectively)
are obtainable from the top equation in their lists (Q4 and H3, respectively)
by wider classes of transformations.  However, it is stressed that these transformations
are outside of the admissible transformations that the classification rests on.
(The issue of transformations of integrable lattice equations was discussed again in
\cite{Hiet:search}, where, again, transformations on sublattices
are observed to sometimes preserve
and sometimes conflict with higher-dimensional consistency.)

Under the ABS symmetry assumptions there are two possible 
forms of the lattice equation,
corresponding to the two possible values of $\sigma$ in equation (\ref{sym2}).

\noindent
$\mathbf{\sigma= 1}$: 
\bea \label{ABSsig1}
Q= a_0 x u v y + a_1(xuv + uvy + vyx + yxu) + a_2(xy + uv) \nonumber\\ + \wb{a}_2(xu+vy) + \wt{a}_2(x v + u y)
+ a_3(x+u+v+y) + a_4. 
\eea

\noindent
$\mathbf{\sigma= -1}$: 
\be \label{ABSsigm1}
Q= a_1(xuv + uvy - vyx - yxu) + a_2(xy - uv) + a_3(x-u-v+y).
\ee
Under the ABS assumptions the $\sigma=-1$ case was shown to be empty.  By removing
the assumption of
the tetrahedron property, Hietarinta was able to derive an integrable lattice
equation with this $\sigma= -1$ form; however, it was later shown to be linearizable
by Ramani, Joshi \textit{et al.} \cite{Ram:deconstruct}.

\section{A change of assumption}\label{AsumptChang}

In this paper, assumption 2 (the symmetry assumption) is modified to 
allow directed edges.  
(A notion of directed edges has appeared 
at least once before in the literature, \cite{BobMeSu:linnonlin},
although
the labelling of the quadrilateral was slightly different to that given here.)
We change assumption 2 to

$\mathbf{2'}$\textbf{ ] Symmetry.}
\\
$Q(x, u, v, y; \alpha, \beta)$ is invariant under the symmetries:

	\be\label{antisymmetry}
		Q(x,u,v,y;\alpha,\beta) = \epsilon Q(x,v,u,y;\beta,\alpha) = \sigma Q(u,x,y,v;-\alpha,\beta)
	\ee
$\epsilon, \sigma = \pm 1$.

The change here extends the reflection symmetry of ABS, $(x,u,v,y) \mapsto (u,x,y,v)$,
in a natural way to include `directed edges', 
in the sense that under this reflection transformation
we also have
$\alpha \mapsto -\alpha$.  This is a generalization of (as opposed to an alternative to)
the ABS classification.  \emph{The ABS classification consists of the set of integrable lattice equations
with these symmetries where the lattice parameters $\{a_i\}$ appear in even functions.}
For example, take the members of the ABS classification as
presented in \cite{AdBoSu:Class}, make the replacement $\alpha_i \mapsto \alpha_i^2$,
for all $i$, and they have this new symmetry.

Under the new symmetry assumption (\ref{antisymmetry}) we can write
\bea
Q= A \,x u v y + B \, xuv + C \, uvy + D \, vyx + E \, yxu 
	+ F \, xy + G \, uv + H \, xu + I \, vy + J \, x v + K \, u y
\nonumber\\ \phantom{abc}+ L \, x + M \, u + N \, v + O \, y + P ,
\label{newsymQ}
\eea
where $A \equiv A(\alpha,\beta),$ etc., and the symmetry conditions
impose the following conditions on the coefficients:
\bee
\begin{array}{rclcrclcrcl}
A(-\alpha,\beta) & \! \! \! = \! \! \! & \sigma A(\alpha,\beta) & , & 
B(-\ab) & \! \! \! = \! \! \! & \sigma E(\ab) & , & 
C(-\ab) & \! \! \! = \! \! \!  & \sigma D(\ab) \\ 
F(-\ab) & \! \! \! = \! \! \!  & \sigma G(\ab) & , & H(-\ab) & \! \! \! = \! \! \!  & \sigma H(\ab) & , &
I(-\ab) & \! \! \! = \! \! \!  & \sigma I(\ab) \\
J(-\ab) & \! \! \! = \! \! \!  & \sigma K(\ab) & , &
L(-\ab) & \! \! \! = \! \! \!  & \sigma M(\ab) & , & N(-\ab) & \! \! \! = \! \! \!  & \sigma O (\ab) \\
P(-\ab) & \! \! \! = \! \! \!  & \sigma P(\ab) & \phantom{,} & 
\phantom{O(\ba)} & \phantom{\! \! \! = \! \! \! } & \phantom{\epsilon O(\ab)} & \phantom{,} & \phantom{P(-\ab)} & \phantom{\! \! \! = \! \! \! } & \phantom{\sigma P(\ab)}
\end{array}
\eee
and
\bee
\begin{array}{rclcrclcrcl}
A(\beta, \alpha) & \! \! \! = \! \! \! & \epsilon A(\ab) & , &
B(\ba) & \! \! \! = \! \! \! & \epsilon B(\ab) &, &  
C(\ba) & \! \! \! = \! \! \! & \epsilon C(\ab) \\
D(\ba) & \! \! \! = \! \! \! & \epsilon E(\ab) & , & 
F(\ba) & \! \! \! = \! \! \! & \epsilon F(\ab) & , & 
G(\ba) & \! \! \! = \! \! \! & \epsilon G(\ab) \\
H(\ba) & \! \! \! = \! \! \! & \epsilon J(\ab) & , & 
I(\ba) & \! \! \! = \! \! \! & \epsilon K(\ab) & , & 
L(\ba) & \! \! \! = \! \! \! & \epsilon L(\ab) \\ 
M(\ba) & \! \! \! = \! \! \! & \epsilon N(\ab) & , & 
O(\ba) & \! \! \! = \! \! \! & \epsilon O(\ab) & , &  
P(\ba) & \! \! \! = \! \! \! & \epsilon P(\ab). 
\end{array}
\eee

To preserve this symmetry the natural transformations under which equations
are identified are now:
\begin{itemize}

	\item Action on all field variables $\{x_i\}$ by one and the 
	same (independent of lattice parameter) M\"obius transformation.
	
	\item Simultaneous point change of all parameters $\alpha \mapsto \phi(\alpha)$ that preserves the parity of 
	the lattice equation coefficients as functions of the parameters.

\end{itemize}
These are our \emph{admissible} transformations.
When the lattice coefficients depend on the lattice parameters
only as even functions (i.e., the lattice equation has
`undirected edges') then the class of 
simultaneous point changes of all parameters is the same as the ABS case; however,
if the lattice coefficients depend on the lattice parameters
in ways other than as even functions, the transformation of parameters must obey $\phi(-\alpha)=-\phi(\alpha)$,
or the reflection symmetry will be lost.
Any simultaneous point change of all parameters will preserve integrability but may break the new symmetry
(so that the lattice equation may have neither symmetry $2$ or $2'$).
Note again that if the lattice equation 
coefficients are even functions of $\alpha$ (undirected edges) 
then the form of the lattice equation,
(\ref{newsymQ}), becomes that of ABS (\ref{ABSsig1}), (\ref{ABSsigm1}).

The fractional multi-linear expressions that arise when solving the lattice equation
(in this case (\ref{newsymQ})) for a particular field variable are an
important ingredient of the algorithmic way of obtaining the Lax matrices from
the equation.  Lattice equations that do not lead to rational expressions (for instance
equations without multi-linear terms) are discounted from the definition of integrability
on account of being trivial.  Lattice equations that do not contain lattice parameters are
also excluded from the definition of integrability (note that in this case the Lax matrices
that follow from the algorithmic construction, guaranteed by three-dimensional
consistency, will not have a spectral parameter).

\section{Derivation: Analysis}\label{Der:a}

We now follow the first part of the classification derivation in \cite{AdBoSu:Class}
(the `analysis') making changes where necessary for the different symmetry condition.

The `analysis' of the classification of ABS proceeds by using the assumptions
to deduce a parameter-less discriminant associated with the quad-graph equation.
Proofs identical to those of \cite{AdBoSu:Class} are omitted.

In \cite{AdBoSu:Class} it is shown that 
\bea
		g(x,x_1;\alpha_1,\alpha_2) \, g(x,x_2;\alpha_2,\alpha_3) \, g(x,x_3;\alpha_3,\alpha_1) \, \nonumber \\
			=	- g(x,x_1;\alpha_1,\alpha_3) \, g(x,x_2;\alpha_2,\alpha_1) \, g(x,x_3;\alpha_3,\alpha_2),\label{gfuncteqn}
\eea
where $g(x,u;\alpha,\beta)$ is a \emph{biquadratic polynomial} in $x$ and $u$, defined by
either of the formulas
\bea
	g(x,u;\alpha,\beta) &=& Q Q_{y v} - Q_y Q_v  \label{gQ}\\
	g(x,v;\beta, \alpha) &=& Q Q_{y u} - Q_y Q_u 
\eea
where $Q = Q(x,u,v,y;\alpha,\beta)$.  We now have a slight change from ABS,
`the polynomial $g$ is symmetric $g(x,u;\alpha,\beta) = g(u,x;\alpha,\beta)$' (\cite{AdBoSu:Class}, page 521, 
equation (19)) 
becomes, due to
the different symmetry,
\be\label{gnewsym}
	g(x,u;\alpha,\beta) = g(u,x;-\alpha,\beta).
\ee
As in \cite{AdBoSu:Class}, one may prove the following lemma.

{\bf Lemma.} \emph{The discriminants of the polynomials $g=g(x,u;\alpha,\beta)$ and 
$\wb{g}= g(x,v;\beta,\alpha)$, considered
as quadratic polynomials in $u$ and $v$, respectively, coincide:
\be\label{discrims}
g_u^2 - 2 g g_{uu} = \wb{g}_v^2 - 2 \wb{g}\wb{g}_{vv}.
\ee}

Equation (\ref{gfuncteqn}) implies properties of the polynomial $g$.
The following proposition is the same as proposition 5 of \cite{AdBoSu:Class}
(the argument of the proof is similar; however, the
differences give us additional information on $k(\alpha, \beta)$).

{\bf Proposition.} \emph{The polynomial $g(x, u; \alpha, \beta)$ may be represented as
\be\label{gkh}
	g(x, u; \alpha, \beta) = k(\alpha,\beta) h (x,u;\alpha),
\ee
where the factor $k$ is antisymmetric,
\be\label{antisymk}
	k(\beta, \alpha) = - k(\alpha,\beta),
\ee
and the coefficients of the polynomial $h(x,u;\alpha)$ depend on the parameter
$\alpha$ in such a way that its discriminant
\be
r(x) = h_u^2 - 2 h h_{uu}
\ee
does not depend on $\alpha$. }

{\bf Proof.}
Equation (\ref{gfuncteqn}) implies, by a separation of variables argument, that
\beq
	\frac{g(x,x_1;\alpha_1,\alpha_2)}{g(x,x_1;\alpha_1,\alpha_3)} =: \frac{f(x;\alpha_1,\alpha_2)}{f(x;\alpha_1,\alpha_3)}
\eeq
and, hence, by the new symmetry $2'$ and a further separation of variables argument
\be\label{gkap}
	\frac{g(x,x_1;\alpha_1,\alpha_2)}{g(x,x_1;\alpha_1,\alpha_3)} = \frac{k(\alpha_1,\alpha_2)}{k(\alpha_1,\alpha_3)},
\ee
as in \cite{AdBoSu:Class}, but with the additional restriction that 
\be\label{krest}
	\frac{k(\alpha_1,\alpha_2)}{k(\alpha_1,\alpha_3)} = \frac{k(-\alpha_1,\alpha_2)}{k(-\alpha_1,\alpha_3)}.
\ee
Then, continuing as in \cite{AdBoSu:Class}, equation (\ref{gfuncteqn}) with (\ref{gkap}) implies the antisymmetry
of $k$, i.e., $k(\alpha,\beta) = - k(\beta,\alpha)$, and (\ref{gkap}) implies (\ref{gkh}).
Equation (\ref{discrims}) implies
\beq
h_u^2-2hh_{uu} = \wb{h}_v^2 - 2 \wb{h}\wb{h}_{vv} \quad , \quad h = h(x,u;\alpha) \quad , \quad \wb{h} = h(x,v;\beta)
\eeq
and therefore $r(x) := h_u^2 - 2h h_{uu}$ does not depend on $\alpha$. $\square$

\smallskip

By a separation of variables argument, equation (\ref{krest}) gives
\be\label{kchi}
\frac{k(\alpha_1,\alpha_2)}{k(-\alpha_1,\alpha_2)} = \frac{\chi(\alpha_1)}{\chi(-\alpha_1)}.
\ee
However, from (\ref{gkap}), $k(\alpha,\beta)$ is only defined up to a factor 
that is a function of $\alpha$.
Hence we may take $k(\alpha,\beta)$ to be defined such that
\be
k(\alpha_1,\alpha_2) = k(-\alpha_1,\alpha_2).
\ee 

The following lemma is new, and is a key fact leading to the derivation of the new lattice equations.

{\bf Lemma.} \emph{The biquadratic $h$ is symmetric in $x$ and $u$,}
\be\label{hsym}
h(x,u,\alpha) = b_0 x^2 u^2 + b_1 (x^2 u + x u^2) + b_3 (x^2 + u^2) + b_5 x u + b_6 (x + u) + b_8,
\ee
\emph{and $\{b_i\}$ are all even functions of $\alpha$.}

{\bf Proof.}
The factorization (\ref{gkh}) in conjunction with the symmetry consequence (\ref{gnewsym}) leads to
\beq
k(\alpha,\beta) \, h(x,u;\alpha) = k(-\alpha,\beta) \, h(u,x;-\alpha),
\eeq
hence
\be
h(x,u;\alpha) = h(u,x;-\alpha).
\ee
If we write the biquadratic $h(x,u;\alpha)$ as
\be
h(x,u,\alpha) = b_0 x^2 u^2 + b_1 x^2 u + b_2 x u^2 + b_3 x^2 + b_4 u^2 + b_5 x u + b_6 x + b_7 u + b_8,
\ee
where $b_i \equiv b_i(\alpha)$, then the coefficients obey the relations:
\beaq
b_1 (-\alpha) & = & b_2(\alpha) \\
b_3 (-\alpha) & = & b_4(\alpha) \\
b_6 (-\alpha) & = & b_7(\alpha) \\
b_0 (-\alpha) & = & b_0(\alpha) \\
b_5 (-\alpha) & = & b_5(\alpha) \\
b_8 (-\alpha) & = & b_8(\alpha). 
\eeaq
Equations (\ref{gkap}) and (\ref{krest}) give
\be\label{gmin}
	\frac{g(x,x_1;\alpha_1,\alpha_2)}{g(x,x_1;\alpha_1,\alpha_3)} = \frac{g(x,x_1;-\alpha_1,\alpha_2)}{g(x,x_1;-\alpha_1,\alpha_3)},
\ee
which implies
\be
g(x,x_1;\alpha_1,\alpha_2) = c \, g(x,x_1;-\alpha_1,\alpha_2).
\ee
The factorization (\ref{gkh}) then leads to
\be
h(x,u;\alpha) = c \, h(x,u;-\alpha).
\ee
Comparing coefficients shows $c=1$ and $h$ is symmetric; that is, $b_2=b_1$, $b_4=b_3$, and $b_7=b_6$.
$\square$

\section{Derivation: Classification extension}\label{der:c}

If the biquadratic $h$, given by (\ref{hsym}), depends on $\alpha$ it does so as an even function.  
However,
if the only dependency on $\alpha$ is as an even function, we may consider
a new dependency on $\alpha' := \alpha^2$.  In terms of $\alpha'$ 
(which is an `undirected edge' lattice parameter)
the new symmetry  
proposed in this paper becomes the reflection symmetry of ABS, and leads to the ABS
classification.  Therefore, if the new symmetry is to lead to additional integrable lattice equations,
augmenting the existing classification, then $h$ must be independent of $\alpha$.

The classification of ABS ranks lattice equations according to canonical forms of the polynomial
$r(x)$ associated with them.  To continue the parallels with the ABS classification (so that
the place in the classification where the new lattice equations belong may easily be seen)
we do the same.

The simultaneous M\"obius transformations 
$x \mapsto (ax+b)/(cx+d)$ transforms the polynomial $r$ as:  
\be
	r(x) \mapsto (cx+d)^4 \, r\left(\frac{ax+b}{cx+d}\right).
\ee
Canonical forms of the quartic polynomial $r(x)$:
\begin{description}
	\item[$r=0$] 
	\item[$r=1$] ($r$ has one quadruple zero);
	\item[$r=x$] ($r$ has one simple zero and one triple zero);
	\item[$r=x^2$] ($r$ has two double zeros);
	\item[$r=x^2-1$] ($r$ has two simple zeros and one double zero);
	\item[$r=4x^3-g_2 x - g_3$, $\Delta = g_2^3 - 27g_3^2 \neq 0$] ($r$ has four simple zeros).
\end{description}

From a given form of $h$, the associated lattice coefficients are found by solving
the nine equations that arise by comparing coefficients of the biquadratic (\ref{gkh})
(the lattice coefficients entering from the definition of $g$, (\ref{gQ}))
and the nine equations obtained by interchanging $\alpha$ and $\beta$ and
using (\ref{antisymk}) and the transformations of the lattice coefficients
given in section \ref{AsumptChang}.  
Considering the canonical forms for $r(x)$ separately gives the following results:

\noindent
{$\mathbf{r=0}$}

The only forms of $h$ (that are independent of $\alpha$) that give $r=0$ and lead to an
integrable lattice equation 
obeying the assumptions of this paper are M\"obius equivalent to $h=1$.

\noindent
{$\mathbf{r=x^2}$}

The only forms of $h$ (that are independent of $\alpha$) that give $r=x^2$ and lead to an
integrable lattice equation 
obeying the assumptions of this paper are M\"obius equivalent to $h= x u$.

\noindent
\textbf{Remaining canonical forms of $\mathbf{r(x)}$}

Although there are biquadratics $h$ independent
of $\alpha$ that reduce
to the particular $r(x)$, there are no associated lattice equations obeying the symmetries given 
in section \ref{AsumptChang}.

\subsection{Lattice equation of new symmetry associated with $r=0$}
Solving 
\beq
g(x, u; \alpha, \beta) = k(\alpha,\beta) h (x,u;\alpha) \equiv k(\alpha,\beta) h (x,u),
\eeq
\beq
g(x, u; \beta, \alpha) = k(\beta, \alpha) h (x,u;\beta) \equiv - k(\alpha,\beta) h (x,u),
\eeq 
with $h(x,u)=1$,
for the lattice coefficients, and imposing the symmetries given 
in section \ref{AsumptChang} leads to
\be\label{Qr0}
Q = K(\alpha,\beta) (x-y)(v-u) + M(\alpha,\beta)(u-v) + \sigma M(-\alpha,\beta) (x-y) + P(\alpha,\beta),
\ee
\be\label{kr0}
k = -P(\alpha,\beta) K(\alpha,\beta) - \sigma M(\alpha,\beta)M(-\alpha,\beta) ,
\ee
and the conditions on the lattice coefficients:
\begin{center}\beq
K(\ba) = -\epsilon K(\ab) \quad , \quad K(-\ab) = \sigma K(\ab), 
\eeq
\beq
M(\ba) = -\epsilon M(\ab) \quad , \quad M(-\ba) = \epsilon M (-\ab), 
\eeq
\beq
P(\ba) = \epsilon P(\ab) \quad , \quad P(-\ab) = \sigma P(\ab).
\eeq
\end{center}
This lattice equation is now a candidate for three-dimensional consistency.

We now slightly re-write the lattice equation, and investigate
the three-dimensional consistency condition, which will give us
the final form of the integrable lattice equation.
Dividing through by $K(\ab)$ and defining $m(\ab) := M(\ab)/K(\ab)$ and
$p(\ab) := P(\ab)/K(\ab)$, gives
\be\label{Qr0mp}
Q = (x-y)(v-u) + m(\alpha,\beta)(u-v) + m(-\alpha,\beta) (x-y) + p(\alpha,\beta),
\ee
and the symmetry conditions become
\begin{center}\beq
m(\ba) = m(\ab) \quad , \quad m(-\ba) = - m (-\ab), 
\eeq
\beq
p(\ba) = - p(\ab) \quad , \quad p(-\ab) = p(\ab).
\eeq
\end{center}
Investigating
three-dimensional consistency gives two conditions that are
necessary and sufficient for it to hold.  The first is
\be\label{3ms}
m(\alpha_3,\alpha_1) + m(-\alpha_1,\alpha_2) - m(\alpha_2,\alpha_3) = 0. 
\ee
Using the symmetry conditions and
differentiating (\ref{3ms}) with respect to $\alpha_3$ gives
\bee
\frac{\partial m(\alpha_3,\alpha_1)}{\partial \alpha_3} =
\frac{\partial m(\alpha_3,\alpha_2)}{\partial \alpha_3} = f'(\alpha_3) , 
\eee
by a separation of variables argument.  Further use of the symmetry conditions yields
\be
m(\alpha_i,\alpha_j) = f(\alpha_i) + f(\alpha_j).
\ee
The symmetry requirement $m(-\ba) = - m (-\ab)$ shows $f(-\alpha) = - f(\alpha)$,
hence, using the admissible transformations, we may take $f(\alpha) = \alpha$.
Therefore $m(\alpha,\beta) = \alpha +\beta$.

The second condition for three-dimensional consistency, after setting 
$m(\alpha,\beta) = \alpha +\beta$, reads
\be\label{3ps}
p(\alpha_3,\alpha_1) + p(\alpha_1,\alpha_2) + p(\alpha_2,\alpha_3) = 0. 
\ee
Differentiating (\ref{3ps}) with respect to $\alpha_2$ and using the symmetry conditions
gives
\be
p(\alpha_1,\alpha_2) = \lambda(\alpha_1) - \lambda(\alpha_2).
\ee
The symmetries show $\lambda(\alpha)$ may be any even function of $\alpha$.
(We may also replace $\lambda(\alpha_i)$ with ${\alpha'}_i^2$ where $\alpha'_i$ is a second
lattice parameter associated with edge $i$.) 
Therefore,
\be\label{Qr0final}
Q = (x-y)(v-u) + (\alpha+\beta)(u-v) + (\beta-\alpha) (x-y) + \lambda(\alpha)-\lambda(\beta).
\ee
We call this lattice equation B1.
(Scaling such that the linear terms, $(\alpha+\beta)(u-v)$ and $(\beta-\alpha)(x-y)$, 
disappear leaves the lattice parameters
only appearing as even functions, hence the lattice equation becomes a member of the ABS classification,
namely H1.)

\subsection{Lattice equation of new symmetry associated with $r=x^2$}
Solving 
\beq
g(x, u; \alpha, \beta) = k(\alpha,\beta) h (x,u;\alpha) \equiv k(\alpha,\beta) h (x,u),
\eeq
\beq
g(x, u; \beta, \alpha) = k(\beta, \alpha) h (x,u;\beta) \equiv - k(\alpha,\beta) h (x,u),
\eeq 
with $h(x,u)=x u$,
for the lattice coefficients, and imposing the symmetries given 
in section \ref{AsumptChang} leads to
\be\label{Qrx2}
Q = \en \sg K(-\ba) x u + \en K(\ba) v y   +  \sg K(-\ab) x v  + K(\ab) u y,
\ee
\be\label{krx2}
k = \sg K(-\ba) K(\ba) - \sg K(-\ab) K(\ab),
\ee
and the condition on the lattice coefficients:
\be\label{rxsqd_sym}
K(\beta,-\alpha) = \sigma K(\ba). 
\ee
This lattice equation is now a candidate for three-dimensional consistency.
Investigating
the three-dimensional consistency condition gives the two conditions:

\be\label{rxsqd_a21}
\frac{K(\alpha_2,\alpha_1)}{K(-\alpha_2,\alpha_1)} = 
\frac{K(\alpha_2,\alpha_3)}{K(-\alpha_2,\alpha_3)}
\ee
(and its cyclic permutations) and
\be\label{rxsqd_a22}
\frac{K(\alpha_1,\alpha_3)}{K(\alpha_3,\alpha_1)} 
\frac{K(\alpha_3,\alpha_2)}{K(\alpha_2,\alpha_3)}
\frac{K(\alpha_2,\alpha_1)}{K(\alpha_1,\alpha_2)}= 
-\epsilon.
\ee
A separation of variables argument from (\ref{rxsqd_a21})
gives
\be\label{rxsqd_f}
\frac{K(\alpha_2,\alpha_1)}{K(-\alpha_2,\alpha_1)} = 
\frac{f(\alpha_2)}{f(-\alpha_2)}.
\ee
Equation (\ref{rxsqd_a22}) implies $\epsilon = -1$ and
\be\label{rxsqd_f}
\frac{K(\alpha_1,\alpha_2)}{K(\alpha_2,\alpha_1)} = 
\frac{\mu(\alpha_1)}{\mu(\alpha_2)}.
\ee
So the symmetry condition (\ref{rxsqd_sym}) may be re-written as
\be\label{rxsqd_symfg}
\sigma \frac{f(\alpha_2)}{f(-\alpha_2)} = \frac{\mu(\alpha_2)}{\mu(-\alpha_2)}. 
\ee
Hence
\be
Q = -\mu(-\beta) \, xu - \mu(\beta) \, vy  + \mu(-\alpha) \, xv + \mu(\alpha) \, uy . 
\ee
If $\mu$ is an even function the lattice equation is part of the ABS classification
(H3$_{\delta=0}$).  If $\mu$ is not an even function we have an integrable lattice equation that
is not part of the ABS classification.  We call this lattice equation B2.

\section{Discussion}\label{discussion}

That B1 and B2 are not part of the ABS classification 
is apparent by noting they are not of the ABS forms
(\ref{ABSsig1}) and (\ref{ABSsigm1}).
B1 is a generalization of the lattice potential KdV.  
Setting $\mu(\alpha) = \alpha$ in B2 it is the 
lattice potential mKdv.  Setting $\mu(\alpha) = -\alpha - r$ in B2 it is the 
`mixed lattice mKdV, discrete-time Toda lattice'. 
See \cite{NiCa:dkdv} for a survey of results
on the lattice potential KdV, lattice potential mKdv,
and `mixed lattice mKdV, discrete-time Toda lattice' up to 1995.
As already discussed, B1 can be reduced to H1 and the form of B2 with 
an even function $\mu$ is H3$_{\delta=0}$.
Further historical remarks on the lattice equations in the ABS classification
can be found in \cite{AdBoSu:Class}.

Using transformations outside of our admissible, symmetry preserving,
transformations, B1 can be transformed to H1 and B2 to H3$_{\delta=0}$;
hence they should not be considered as new integrable lattice equations.
(In the same way that, for instance, A1 can be transformed to
Q1.)
Rather, the main result is that these are the only two
integrable lattice equations with the given symmetry assumptions and directed edges.

\section{Conclusion}

The ABS classification, \cite{AdBoSu:Class}, has been extended
by allowing the natural notion of directed edges.
The full classification is shown in table \ref{Newclass}.

\begin{table}
\begin{center}
\begin{tabular}{ | c | c | c | c | c |} \hline
Discriminant $r(x)$ & ABS: Q list & ABS: A list & ABS: H list & New symmetry \\ \hline 
$r(x)=0$ & Q1${}_{\delta=0}$ & A1${}_{\delta=0}$ & H1 & B1 \\\hline
$r(x)=1$ & Q1$_{\delta=1}$ & A1$_{\delta=1}$ & H2 & {} \\\hline 
$r(x)=x$ & Q2 & {} & {} & {} \\ \hline
$r(x)=x^2$ & Q3$_{\delta=0}$ & A2 & H3 & B2 \\ \hline
$r(x)=x^2-\delta^2$ & Q3 & {} & {} & {} \\ \hline
$r(x)=4x^3 - g_2 x - g_3$ & Q4 & {} & {} & {} \\ \hline
\end{tabular}
\caption{\label{Newclass}The classification of integrable lattice equations with symmetries $1$, $2'$ and $3$.}
\end{center}
\end{table}

Our admissible transformations include simultaneous point change of all 
parameters that preserve the parity of the 
coefficients of the lattice equations as functions of the lattice parameters.  
This preserves the symmetry of the lattice equation.  However,
any simultaneous point change of all 
parameters preserves the integrability, but destroys the symmetry.  
This indicates that symmetry stipulations
lose their importance on a deeper level. 
Table \ref{Newclass} is currently the largest list of
integrable lattice equations; however, it would be 
desirable to have the full classification 
(or even the full list) of integrable lattice equations
without symmetry assumptions.  This remains an open problem.

    \section{Acknowledgements}
The author wishes to thank Nalini Joshi for numerous discussions on integrable lattice equations,
and for
advice and encouragement.  
He wishes to thank Chris Ormerod for proofreading this paper
and Jarmo Hietarinta for useful correspondence on
transformations of integrable lattice equations.
The author is supported by
the Australian Research Council Discovery Project Grant \#DP0664624.

\end{document}